\documentclass[showpacs,preprintnumbers,amsmath,amssymb,twocolumn,superscriptaddress,prb]{revtex4}
\usepackage{times}
\usepackage{graphicx}
\usepackage{amsfonts}
\usepackage{amsmath, amsthm, amssymb}
\usepackage{dsfont}
\usepackage{subfigure}
\newcommand{\vp}{\mathbf{p}} 
\newcommand{\vpf}{\mathbf{p}_\text{F}}
\newcommand{\vecr}{\mathbf{R}}
 
\newcommand{\e}[1]{\mathrm{e}^{#1}}
\newcommand{\g}{\check{g}(\vpf,\vecr;\varepsilon,t)}
\newcommand{\G}{\check{G}(\vp,\vecr;\varepsilon,t)} 
\newcommand{\ug}{\underline{g}}
\newcommand{\uf}{\underline{f}}

\newcommand{\lessim}{\mbox{\tiny$\mbox{\normalsize$<$}\atop
\mbox{\normalsize$\sim$}$}}
\newcommand{\grtsim}{\mbox{\tiny$\mbox{\normalsize$>$}\atop
\mbox{\normalsize$\sim$}$}}

\newcommand{\eg}{\textit{e.g. }}%[syn: f.eks., for example, for instance]
\newcommand{\etal}{\emph{et al. }}
\def\i{\mathrm{i}}

\begin{document}
\title[The role of interface transparency and spin-dependent scattering in diffusive ferromagnet/superconductor heterostructures]{The role of interface transparency and spin-dependent scattering in diffusive ferromagnet/superconductor heterostructures}
\author{Jacob Linder}
\affiliation{Department of Physics, Norwegian University of
Science and Technology, N-7491 Trondheim, Norway}
\author{Takehito Yokoyama}
\affiliation{Department of Applied Physics, Nagoya University, Nagoya, 464-8603, Japan}
\author{Asle Sudb{\o}}
\affiliation{Department of Physics, Norwegian University of
Science and Technology, N-7491 Trondheim, Norway}

\date{Received \today}
\begin{abstract}
We present a numerical study of the density of states in a ferromagnet/superconductor junction and the Josephson current in a superconductor/ferromagnet/superconductor junction in the diffusive limit by solving the Usadel equation with Nazarov's boundary conditions. Our calculations are valid for an arbitrary interface transparency and arbitrary spin-dependent scattering rate, which allows us to explore the entire proximity-effect regime. We first investigate how the proximity-induced anomalous Green's function affects the density of states in the ferromagnet for three magnitudes of the exchange field $h$ compared to the superconducting gap $\Delta$: \textit{i)} $h\lessim\Delta$, \textit{ii)} $h\gtrsim \Delta$, \textit{iii)} $h\gg \Delta$. In each case, we consider the effect of the barrier transparency and allow for various concentrations of magnetic impurities. We clarify features that may be expected in the various parameter regimes accessible for the ferromagnetic film, with regard to thickness and exchange field. In particular, we address how the zero-energy peak and minigap observed in experiments may be understood in terms of the interplay between the singlet and triplet anomalous Green's function and their dependence on the concentration of magnetic impurities. Our results should serve as a useful tool for quantitative analysis of experimental data. We also investigate the role of the barrier transparency and spin-flip scattering in a superconductor/ferromagnet/superconductor junction. We suggest that such diffusive Josephson junctions with large residual values of the supercurrent at the 0-$\pi$ transition, where the first harmonic term in the current vanishes, may be used as efficient supercurrent-switching devices. We numerically solve for the Josephson current in such a junction to clarify to what extent this idea may be realized in an experimental setup. It is also found that uniaxial spin-flip scattering has very different effect on the 0-$\pi$ transition points depending on whether one regards the width- or temperature-dependence of the current. Our theory takes into account vital elements that are necessary to obtain quantitative predictions of the supercurrent in such junctions.
\end{abstract}
\pacs{74.20.Rp, 74.50.+r,74.70.Kn}

\maketitle

\section{Introduction}
Proximity structures consisting of ferromagnetic and superconducting materials are nowadays a very active research field. The interest in this 
type of systems has grown considerably over the last decade, since they offer novel and interesting phenomena to explore from a fundamental 
physics point of view. In addition, it is hoped that future applications in low-temperature nanotechnology may emerge from this research field. Ferromagnetism is usually considered to be antagonistic to conventional superconductors, since the exchange field acts as a depairing agent for spin-singlet Cooper pairs. However, closer examination reveals that the physical situation is more subtle than that. The proximity effect on a superconductor from a ferromagnet does not merely suppress the spin-singlet superconducting order parameter, but may also under specific circumstances induce exotic features such as odd-frequency pairing and long-ranged spin-triplet correlations \cite{bergeretRMP, buzdin}.
\par
Various theoretical idealizations allow for a relatively simple approach to ferromagnet/superconductor (F/S) heterostructures in the quasiclassical framework. One 
of the most popular approaches in the literature employs the linearized Usadel \cite{usadel} equations with the Kupriyanov-Lukichev boundary 
conditions \cite{kupluk}, which is a viable method in the case of a weak proximity effect. This is for instance permissable when the 
barrier transparency of the F/S interface is low. Although the linearized treatment clearly represents a special limit, much useful 
information has been obtained through this approach. 
 \begin{figure}[h!]
\centering
\resizebox{0.45\textwidth}{!}{
\includegraphics{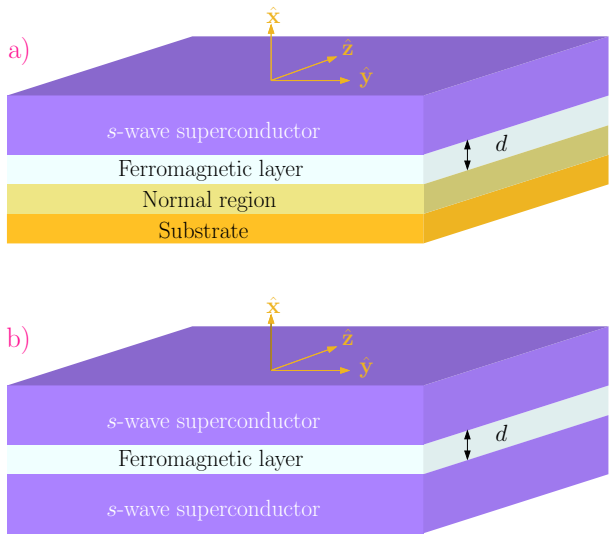}}
\caption{(Color online). The ferromagnet/superconductor heterostructures investigated in this paper. In a), a ferromagnet/superconductor bilayer setup is shown. We will model the barrier region at the normal/ferromagnet
interface to be very strong, mimicking a free edge boundary condition or insulating region. Our proposed experimental setup is thus equivalent to the ones employed in the experimental works of Refs.~\onlinecite{kontos, beasley2}. In b), we consider a superconductor/ferromagnet/superconductor junction. The superconductors are treated as reservoirs, thus unaffected by the proximity effect. We will consider a ferromagnetic layer thickness ranging from $d/\xi=0.1$ to $d/\xi=1.0$, where $\xi$ is the superconducting coherence length.}
\label{fig:model}
\end{figure}
Among these are the oscillations of the anomalous Green's function 
\cite{lazar,buzdinbrief} in the ferromagnet due to the fact that the Cooper pair in the ferromagnet acquires a finite center-of-mass 
momentum $q=2h/v_\text{F}$, where $h$ is the magnetic exchange energy and $v_\text{F}$ is the Fermi velocity. \cite{Demler} Although several other 
works have also considered various aspects of the density of states in both diffusive \cite{Baladie,Golubov2002,Krivoruchko,bergeret01, gosu1, audrey, linder07,Cottet} and clean \cite{Pilgram,zareyan,kopu} ferromagnet/superconductor junctions, most of these rely on simplifying assumptions concerning the interface, 
 that it is either perfectly transparent or strongly insulating.
\par
From an experimental point of view, there has been several investigations of how the spin-splitting in energy level in a ferromagnet affects the proximity 
effect when placed in contact with superconductor \cite{ryazanov, kontos,beasley1,Cretinon}. Very recently, SanGiorgio \etal \cite{beasley2} 
reported an anomalous double peak structure in the density of states of a Nb/Ni tunneling junction. They attempted to capture the 
qualitative features of the unusual subgap-structure of the density of states by using the Usadel equation, but were not able to do 
so. This clearly warrants further investigation, and serves as a motivation for employing more sophisticated models of a 
ferromagnet/superconductor interface, possibly including a domain structure in the ferromagnet.
\par
Recently, a full numerical solution of the Usadel equation was employed by Gusakova \etal including scattering on magnetic impurities, but under 
the simplifying assumption that the tunneling limit was reached at the F/S interface \cite{gosu2}. It would clearly also be of interest to consider higher barrier transparencies to elucidate how this may influence the proximity effect. In this case, one should employ Nazarov's boundary conditions \cite{nazarov} instead of the Kupriyanov-Lukichev boundary conditions \cite{kupluk}. Yokoyama \etal recently adopted this approach for the case without any magnetic impurities present \cite{yokoyama,yokoyama07}. When comparing the theoretical models against quantitative aspects of the experimental data, the influence of magnetic impurities and barrier transparency clearly play a pivotal role. For instance, as argued in Ref.~\onlinecite{zareyan2}, the predicted amplitude of the critical current in diffusive S/F/S junction is $\sim 10^3$ times larger than the actual measured curves. This may be attributed to the strong suppression of Andreev bound states due to spin-flip scattering processes, which are usually not taken into account in the theoretical treatment (see, however, Ref.~\onlinecite{bergeretjos}). To date, there exists no study of the density of states in F/S structures that allows access to both the full range of barrier transparencies \textit{and} concentration of magnetic impurities.
\par
In the present paper, we remedy this by employing a numerical solution of the Usadel equation in the ferromagnetic region, taking into account an arbitrary concentration of magnetic impurities as well as arbitrary barrier transparency. This permits us to comprehensively study the local density of states (DOS) in a dirty ferromagnet/superconductor junction for experimentally realistic parameters. We study three cases for the size of the exchange field $h$ compared to the superconducting gap $\Delta$: \textit{i)} $h\lessim\Delta$, \textit{ii)} $h\gtrsim \Delta$, \textit{iii)} $h\gg \Delta$. In each case, we compare the short-junction regime to the wide-junction regime to obtain the energy-resolved DOS. Among other things, we study the emergence of a zero-energy peak in the spectrum in a certain parameter range in addition to the transition from minigaps to peaks in the spectrum for increasing exchange field in the wide-junction regime. In particular, we investigate the interplay between the proximity-induced singlet and triplet anomalous Green's function in the ferromagnet and how these are affected by spin-dependent scattering. Since our results take into account an arbitrary proximity effect and magnetic impurity concentration, they should serve as a useful tool for performing quantitative analysis of experimental data. We envision a ferromagnet/superconductor bilayer as shown in Fig. \ref{fig:model}, which is virtually identical to the experimental setup used in Refs.~\onlinecite{kontos, beasley2}. 
\par
Another interesting issue in the context of ferromagnet-superconductor is that there have been put forth suggestions of exploiting superconductor/ferromagnet/superconductor (S/F/S) Josephson junctions as qubits in quantum computers. By now, the 0-$\pi$ oscillations that occur in S/F/S junctions are also well-established both theoretically \cite{pit1,pit2,pit4,belzigJOS,cayssol,Golubov,linderPRL} and experimentally \cite{ryazanov, pie1, ryazanovprb, kontos,guichard,sellier1,sellier2}. 
\par
As the fundamental understanding of the physics in an S/F/S junction begins to take shape, the next step to take should be a more sophisticated modelling of such systems in order to achieve better quantitative agreement between theory and experiment. Parameters such as barrier transparency and magnetic impurities become important in this respect, as they can have crucial impact on the behaviour of the supercurrent in an S/F/S system. To this end, several recent works investigated some aspects of the supercurrent by including magnetic impurities \cite{bergeretjos,Buzdin2,Houzet,golubovjos,Faure,Kashuba,Mori}, but restricting themselves to the tunneling limit. Under the opposite simplifying assumption of transparent interfaces, the influence of spin-flip scattering has also been investigated \cite{oboznov}. The Josephson current in an S/F/S junction may under these circumstances be written to good accuracy as $I=I_0\sin\phi$, where $\phi$ is the macroscopic phase difference. As a direct result, the usual 0-$\pi$ oscillations observed in S/F/S junctions do not exhibit any residual value of the supercurrent at the transition points where the current switches sign. 
\par
However, an interesting opportunity presents itself in the case when the current-phase relationship deviates strongly from its usual sinusoidal behavior. Higher harmonics in the Josephson current will lead to a finite residual value right at the cusps of the critical current oscillations with exchange field or temperature, and these cusps are indicative of a sign reversal of the supercurrent. This simple observation gives rise to a very interesting prospect. \textit{If} the residual value of the current at the cusps were to be of considerable magnitude, one could exploit this effect to create a dissipationless current \textit{switching} device. In this case, the direction of the Josephson current could be instantaneously flipped by some external control parameter when the system parameters are such that the junction is very close to the 0-$\pi$ transition point. 
\par
Such an idea depends, however, on the possibility to actually obtain a large residual value of the current near the cusps. The early experiments \cite{ryazanov, ryazanovprb, kontos, guichard, sellier1} measuring the Josephson current in diffusive S/F/S systems reported that the critical current vanished at these cusps. But recently, Sellier \etal \cite{sellier2} observed a finite albeit small supercurrent at the 0-$\pi$ transition point where the first harmonical term vanishes.
\par
In Ref. \onlinecite{zareyan2}, the magnitude of the residual value of the current at the cusps was investigated by solving the linearized Usadel equation \cite{usadel} using Nazarov's boundary conditions \cite{nazarov}. However, this calculation was performed assuming two limiting circumstances: \textit{i)} no magnetic impurities and \textit{ii)} a weak proximity effect. The residual value was also analyzed in the clean limit in Ref.~\onlinecite{bobkova}. It is definitely of much interest to go beyond these approximations and see how the physical picture is altered. As mentioned previously, a consequence of \textit{i)} is that the theoretically predicted magnitude of the critical current is a factor $10^3$ larger than the experimentally measured value. Even though highly valuable qualitative information may be obtained in approximations such as \textit{i)} and \textit{ii)}, the progress on both the theoretical and experimental side of S/F/S Josephson junctions calls for a higher accuracy of the quantitative predictions. Obviously, this is also of paramount importance in the context of discussing a practical supercurrent-switch device.
\par
We here develop a theory for the supercurrent in an S/F/S junction which takes into account an arbitrary concentration of magnetic impurities and arbitrary transparency of the interfaces. Both of these are of vital importance in obtaining a quantitative agreement with experimental findings. We numerically solve the Usadel equation and employ Nazarov's boundary condition. In the intermediate barrier transparency regime, we find that a finite but small residual value of the current is permitted at the 0-$\pi$ transition points. However, the effect of spin-flip scattering is very different when comparing the width-dependence against the temperature-dependence of the current. The transition point is much more robust in the width-dependence, although increasing the concentration of magnetic impurities reduces the residual value. On the other hand, increasing spin-flip scattering completely removes the 0-$\pi$ transition point for the temperature-dependence. 
\par
This paper is organized as follows. In Sec. \ref{sec:theory}, we establish the theoretical framework which will serve as our tool to obtain both the density of states in the diffusive ferromagnet/superconductor bilayer and the Josephson current in the diffusive superconductor/ferromagnet/superconductor junction. In Sec. \ref{sec:results}, we present our results with a discussion of these 
in Secs. \ref{sec:anomalous} and \ref{sec:minigap}, as well as
in Sec. \ref{sec:discuss}. Finally, we summarize the results in Sec. \ref{sec:summary}.
\par
\section{Theoretical formulation}\label{sec:theory}
The central quantity in the quasiclassical theory of superconductivity is the quasiclassical Green's functions 
$\g$, which depends on the momentum at Fermi level $\vpf$, the spatial coordinate $\vecr$, energy measured 
from the chemical potential $\varepsilon$, and time $t$. A considerable literature covers the Keldysh formalism 
and non-equilibrium Green's functions \cite{serene, kopnin, rammer, zagoskin,Chandrasekhar,jpdiplom}. Here we only briefly sketch the 
theoretical structure, for the sake of readability and for establishing notation. The quasiclassical Green's 
functions $\g$ is obtained from the Gor'kov Green's functions $\G$ by integrating out the dependence on kinetic 
energy, assuming that $\check{G}$ is strongly peaked at Fermi level,
\begin{equation}\label{eq:quasiclassical}
\g = \frac{\i}{\pi} \int \text{d}\xi_\vp \G.
\end{equation}
The above assumption is typically applicable to superconducting systems where the characteristic length scale of the perturbations 
present, namely superconducting coherence length, is much larger than the Fermi wavelength. The 
corresponding characteristic energies of such phenomena must be much smaller than the Fermi energy $\varepsilon_\text{F}$. 
The quasiclassical Green's functions may be divided into an advanced (A), retarded (R), and Keldysh (K) component, each 
of which has a $4\times4$ matrix structure in the combined particle-hole and spin space.  One has that
\begin{equation}
\check{g} = \begin{pmatrix}
\hat{g}^\text{R} & \hat{g}^\text{K}\\
0 & \hat{g}^\text{A} \\
\end{pmatrix},
\end{equation}
where the elements of $\g$ read
\begin{equation}
\hat{g}^\text{R,A} = \begin{pmatrix}
\ug^\text{R,A} & \uf^\text{R,A}\\
-\tilde{\uf}^\text{R,A} & -\tilde{\ug}^\text{R,A} \\
\end{pmatrix},\;
\hat{g}^\text{K} = \begin{pmatrix}
\ug^\text{K} & \uf^\text{K}\\
\tilde{\uf}^\text{K} & \tilde{\ug}^\text{K} \\
\end{pmatrix}.
\end{equation}
The quantities $\ug$ and $\uf$ are $2\times2$ spin matrices, with the structure
\begin{equation}
\ug = \begin{pmatrix}
g_{\uparrow\uparrow} & g_{\uparrow\downarrow} \\
g_{\downarrow\uparrow} & g_{\downarrow\downarrow} \\
\end{pmatrix}.
\end{equation}
Due to internal symmetry relations between these Green's functions, all of these quantities are not independent. In particular, the 
tilde-operation is defined as
\begin{equation}
\tilde{f}(\vpf,\vecr;\varepsilon,t) = f(-\vpf,\vecr;-\varepsilon,t)^*.
\end{equation}
The quasiclassical Green's functions $\g$ may be determined by solving the Eilenberger \cite{eilenberger} equation
\begin{equation}\label{eq:eilenberger}
[\varepsilon\hat{\rho}_3 - \hat{\Sigma}, \check{g}]_\otimes + \i\mathbf{v}_\text{F}\boldsymbol{\nabla} \check{g} = 0,
\end{equation}
where $\hat{\Sigma}$ contains the self-energies in the system such as impurity scattering, superconducting order parameter, and exchange 
fields. The star-product $\otimes$ is noncommutative and is defined in Appendix. When there is no explicit time-dependence in the problem, 
the star-product reduces to normal multiplication. This is the case we will consider throughout the paper. The operation $\hat{\rho}_3\check{g}$ inside the commutator should be understood as $\hat{\rho}_3\check{g} \equiv \text{diag}\{ \hat{\rho}_3,\hat{\rho}_3\} \check{g}$.
Pauli-matrices in particle-hole$\times$spin (Nambu) space are denoted as $\hat{\rho}_i$, while Pauli-matrices in spin-space are 
written as $\underline{\tau}_i$, all of which are defined in the Appendix. The Green's functions also satisfy the normalization condition 
\begin{equation}
\check{g}\otimes\check{g} = \check{1}.
\end{equation}
In the special case of an equilibrium situation, one may express the Keldysh component in terms of the retarded and advanced Green's 
function by means of the relation
\begin{equation}\label{eq:kra}
\hat{g}^\text{K} = (\hat{g}^\text{R} - \hat{g}^\text{A})\tanh(\beta\varepsilon/2),
\end{equation}
where $\beta = T^{-1}$ is inverse temperature. In nonequilibrium situations, one must derive kinetic equations for nonequilbrium 
distribution functions in order to specify the Keldysh part \cite{belzigreview}.
\par
The above equations suffice to completely describe for instance a single superconducting structure, but must be supplemented by boundary 
conditions when treating heterostructures such as F/S junctions. These boundary conditions take different forms depending on the physical 
properties of the interface. The Kupriyanov-Lukichev \cite{kupluk} boundary conditions may be applied for a dirty junction in the tunneling 
limit when the transparency of the interface is low, while an arbitrary interface transparency requires usage of the boundary conditions 
developed by Nazarov \cite{nazarov}. Boundary conditions for a spin-active interface have also been derived \cite{hh,millis}.
\par
We will consider the dirty limit of the Eilenberger equation Eq. (\ref{eq:eilenberger}), which leads to the Usadel equation \cite{usadel}. Our motivation is that this is the experimentally most relevant situation. This will be an appropriate starting point for diffusive systems where the scattering time due to impurities satisfies $X\tau \ll 1$, where $X$ is the energy scale of any other self-energy in the problem. For strong ferromagnets where $h$ becomes comparable to $\varepsilon_\text{F}$, the stated inequality may strictly speaking not be valid for $X=h$. Hence, 
we will restrict ourselves to the regime $h/\varepsilon_\text{F} \ll 1$ and assume that \cite{pit4} $h\tau \ll 1$. Below, we will mostly concern ourselves with the retarded part of $\g$, 
since the advanced component may be found via the relation
\begin{equation}\label{eq:gagr}
\hat{g}^\text{A} = -(\hat{\rho}_3 \hat{g}^\text{R} \hat{\rho}_3)^\dag.
\end{equation}
By isotropizing the Green's function due to the assumed frequent impurity scattering, it is rendered independent of $\vpf$. This isotropic 
(in momentum space) Green's function satisfies the Usadel equation in the ferromagnet:
\begin{equation}\label{eq:usadel1}
D\nabla (\check{g}\nabla\check{g}) + \i[\varepsilon\hat{\rho}_3 + \hat{M} - \check{\sigma}_\text{sf} -  \check{\sigma}_\text{so} , \check{g}] = 0.
\end{equation}
Above, the exchange energy $h$ is accounted for by the matrix $\hat{M} = \text{diag}(h\underline{\tau_3},h\underline{\tau_3})$, assuming a magnetization in the $\mathbf{z}$-direction, while the spin-flip self-energy reads
\begin{align}
\check{\sigma}_\text{sf}(\vecr;\varepsilon) &= -\frac{\i}{8\tau_\text{sf}} \sum_i \hat{\alpha}_i \check{g}(\vecr;\varepsilon)\hat{\alpha}_iS_i,\notag\\
\check{\sigma}_\text{so}(\vecr;\varepsilon) &= -\frac{\i}{8\tau_\text{so}} \sum_i \hat{\alpha}_i \hat{\rho}_3 \check{g}(\vecr;\varepsilon)\hat{\rho}_3\hat{\alpha}_i,
\end{align}
where $\tau_\text{sf}$ is the spin-flip scattering time and $S_i$ is a spin expectation value, while $\tau_\text{so}$ is the spin-orbit scattering time. We have defined the matrices $\hat{\alpha}_i = \text{diag}(
\underline{\tau_i},\underline{\tau_i}^\text{T})$. The diffusion constant is given by $D=v_\text{F}^2\tau_\text{imp}/3$, where $\tau_\text{imp}$ is the impurity scattering relaxation time. We will here consider either uniaxial spin-flip scattering, such that $S_3=1$ ($\hat{\mathbf{z}}$-direction) and zero otherwise, and also isotropic 
spin-flip scattering where $S_i=1$ for $i\in\{1,2,3\}$. For later use, we denote $S_{xy}\equiv S_1=S_2$ and $S_z=S_3$.
\par
Let us now consider the retarded part of Eq. (\ref{eq:usadel1}) which has the same form, 
namely
\begin{equation}\label{eq:usadelR}
D\nabla (\hat{g}^\text{R}\nabla\hat{g}^\text{R}) + \i[\varepsilon\hat{\tau}_3 + \hat{M} - \hat{\sigma}_\text{sf} - \hat{\sigma}_\text{so}, \hat{g}^\text{R}] = 0.
\end{equation}
From now on, we will omit the superscript 'R' on the Green's function. Moreover, we will find it convenient to parametrize the Green's function 
by exploiting the normalization condition. In the $s$-wave superconductor $(x>d)$ and normal metal $(x<0)$, we use the bulk solutions
\begin{equation}\label{eq:bulk}
\hat{g}_S = \begin{pmatrix}
g\underline{1} & f\i\underline{\tau_2}\\
f\i\underline{\tau_2} & -g\underline{1}\\
\end{pmatrix},\;
\hat{g}_N = \begin{pmatrix}
\underline{1} & \underline{0}\\
\underline{0} & -\underline{1}\\
\end{pmatrix},
\end{equation}
where $g\equiv \text{cosh}(\theta_s)$, $f\equiv \text{sinh}(\theta_s)$, $\theta_s = \text{atanh}(\Delta/\varepsilon)$. 
The Green's function in the ferromagnet may conveniently be parametrized as \cite{jpdiplom}
\begin{equation}\label{eq:gf}
\hat{g}_F = \begin{pmatrix}
\cosh\theta_\uparrow(\varepsilon) & 0 & 0 & \sinh\theta_\uparrow(\varepsilon)  \\
0 & \cosh\theta_\downarrow(\varepsilon) & \sinh\theta_\downarrow(\varepsilon)  & 0 \\
0 & -\sinh\theta_\downarrow(\varepsilon)  & -\cosh\theta_\downarrow(\varepsilon)  & 0 \\
-\sinh\theta_\uparrow(\varepsilon)  & 0 &0 & -\cosh\theta_\uparrow(\varepsilon) \\
\end{pmatrix},
\end{equation}
which satisfies $\hat{g}_F^2 = \hat{1}$. 
We have made use of the symmetry 
$\theta_\uparrow(\varepsilon) = \theta_\downarrow^*(-\varepsilon)$  in obtaining Eq. (\ref{eq:gf}). Note that for $h=0$, 
$\theta_\uparrow(\varepsilon) = -\theta_\downarrow(\varepsilon)$ is satisfied. Note that in Eq. (\ref{eq:gf}), there is both a singlet and triplet component of the anomalous Green's function. The triplet component is opposite-spin paired $(S_z=0)$, and there are no equal-spin pairing $(S_z=\pm1)$ components in our system since we consider homogeneous magnetization and a spin-inactive barrier. The $S_z=0$ triplet component nevertheless plays a pivotal role in interpreting the behaviour of the density of states, as we shall see later. This is because it has a special symmetry property refered to as odd-in-frequency, which will be elaborated upon in Sec. \ref{sec:zep}.
\par
The Usadel equation Eq. (\ref{eq:usadelR}) then yields for majority and minority spin ($\sigma=\uparrow,\downarrow=\pm 1$) in the ferromagnet:
\begin{align}\label{eq:usa}
D\partial_x^2\theta_\sigma& + 2\i(\varepsilon+\sigma h)\sinh\theta_\sigma - \frac{\sigma S_{xy}}{2\tau_\text{sf}}\sinh(\theta_\uparrow-\theta_\downarrow) \notag\\
&- \frac{S_z}{4\tau_\text{sf}}\sinh2\theta_\sigma - \frac{1}{2\tau_\text{so}} \sinh(\theta_\uparrow+\theta_\downarrow) = 0.
\end{align}
We now use the generalized Nazarov boundary condition valid for the diffusive regime \cite{nazarov}. At $x=\{0,d\}$, it reads
\begin{equation}\label{eq:bound}
2\gamma_{L,R} L \hat{g}_F\partial_x \hat{g}_F\Bigg|_{x=\{0,d\}} = \mp \frac{4\tau [\hat{g}_F,\hat{g}_{L,R}]}{4 + \tau(\{\hat{g}_F,\hat{g}_{L,R}\}-2)} \Bigg|_{x=\{0,d\}},
\end{equation}
where $[\ldots]$ and $\{\ldots\}$ denote the commutator and anticommutator, respectively, and $\hat{g}_{L,R}(=\hat{g}_{N,S})$ denotes the Green's function on 
the left and right side of the ferromagnet. The parameter $\gamma_{L,R}=R_B^{L,R}/R_F$ denotes the ratio between the resistance in the left/right 
barrier region $R_B^{L,R}$ and the resistance in the F region $R_F$. We have conventionally introduced the parameter $\tau$, which is the transmissivity of the interface \cite{nazarov}. Giving an expression for $\tau$ in terms of microscopic parameters of the interface 
is not very practical, and we will therefore use $\tau$ as a phenomenological parameter to characterize the transparency of the interface, 
in complete analogy with Eq. 36 of Ref. \onlinecite{nazarov}.  Here, $\tau=0$ corresponds to zero transmission 
of quasiparticles incident on the superconducting interface, and $\tau=1$ corresponds 
to perfect transmission. Below, we will consider two values of $\tau$, namely $0.1$ corresponding to low 
transmissivity, and $\tau=0.5$ corresponding to intermediate transmissivity. Note that the parameters $\tau$ and $\gamma$ may be varied 
independently \cite{tanakaPRB03}, and one does not in general have $\gamma \sim \tau^{-1}$. The reason for this is that $\gamma$ is 
related to the constriction area of the junction, which may be altered independently of the scattering strength proportional to $\tau^{-1}$. 
Inserting Eqs. (\ref{eq:bulk}) and (\ref{eq:gf}) into Eq. (\ref{eq:bound}) yields the
boundary conditions
\begin{align}\label{eq:majbound}
\gamma_L d \partial_x\theta_\uparrow &= \frac{2\tau\sinh\theta_\uparrow}{2-\tau + \tau\cosh\theta_\uparrow},\notag\\
\gamma_L d \partial_x \theta_\downarrow &= \frac{2\tau\sinh\theta_\downarrow}{2-\tau + \tau\cosh\theta_\downarrow},
\end{align}
at the normal/ferromagnet (N/F) interface $(x=0)$, while at the F/S interface $(x=d)$ we have
\begin{align}\label{eq:minbound}
\gamma_R d \partial_x\theta_\uparrow &= \frac{2\tau(\cosh\theta_\uparrow f - \sinh\theta_\uparrow g)}{2-\tau + \tau(\cosh\theta_\uparrow g - \sinh\theta_\uparrow f)},\notag\\
\gamma_R d \partial_x\theta_\downarrow &= \frac{-2\tau(\cosh\theta_\downarrow f + \sinh\theta_\downarrow g)}{2-\tau + \tau(\cosh\theta_\downarrow g + \sinh\theta_\downarrow f)}.
\end{align}
For later use, we define the Thouless energy $\varepsilon_T = D/d^2$. As a measure of the strength of the spin-flip and spin-orbit scattering, which increases with decreasing spin relaxation time, we introduce $g_\text{sf} = \tau_\text{sf}^{-1}$ and $g_\text{so} = \tau_\text{so}^{-1}$. Also note that the $S_z=S_{xy}=1$ for isotropic spin-flip scattering, while $S_z=3$ and $S_{xy}=0$ for uniaxial spin-flip scattering along the $\mathbf{z}$-direction.
The spin-resolved and normalized DOS is obtained as
\begin{align}
N_\uparrow = \text{Re}\{\cosh\theta_\uparrow\},\; N_\downarrow = \text{Re}\{\cosh\theta_\downarrow\}.
\end{align}
Furthermore, we define the total DOS as $N = \sum_\sigma N_\sigma/2$. Eqs. (\ref{eq:usa}), (\ref{eq:majbound}), and (\ref{eq:minbound}) now constitute two coupled nonlinear second order differential equations supplemented 
with boundary conditions which may be solved numerically.
\par
Before presenting our results for the DOS, we establish the theoretical framework for our treatment of the Josephson current in an S/F/S junction with spin-dependent scattering, following the notation of Ref.~\onlinecite{yokoyamaj}. The physical system studied here consists of a junction with two $s$-wave superconductors separated by a diffusive ferromagnet with a resistance $R_F$ and length $d$ much larger than the mean free path. We here only consider uniaxial spin-flip scattering. The interface regions are characterized by a resistance $R_B$. The transparencies of the junction interfaces are given by $T = 4\cos^2\varphi/(4\cos^2\varphi + Z^2)$ where $Z$ is a measure of the barrier strength, and the barriers themselves are considered to be spin-inactive and modelled by infinitely narrow insulating barriers $U(x) = (Zv_F/2)[\delta(x-d) + \delta(x)]$. Above, $v_F$ is the Fermi velocity and $\varphi$ is the injection angle measured from the interface normal to the junction. We employ the quasiclassical theory of superconductivity \cite{serene, kopnin, rammer, zagoskin,Chandrasekhar,jpdiplom}, and make use of the $\theta$-parametrization \cite{belzigreview} of the Green's function. The retarded part $\hat{g}^\text{R}$ may, due to symmetry requirements, be written as \cite{yokoyamaj}
\begin{align}
\ug = \sin\theta(\cos\psi\underline{\tau_1} + \sin\psi\underline{\tau_2}) + \cos\theta\underline{\tau_3}.
\end{align}
From the Usadel quation, one obtains
\begin{align}\label{eq:usadel}
D[\partial_x^2\theta - (\partial_x\psi)^2\cos\theta & \sin\theta] + 2\i(\varepsilon +\sigma h + \i\gamma)\sin\theta = 0,\notag\\
&\partial_x[\sin^2\theta(\partial_x\psi)] = 0,
\end{align}
for $\sigma=\uparrow,\downarrow$ spins. Here, $D$ is the diffusion constant, $h$ is the exchange field, and $\gamma$ is the self-energy associated with uniaxial spin-flip scattering. We employ the bulk solutions of the Green's function in the superconducting regions, assuming that these are much less disordered than the ferromagnetic layer. To gain access to the full regime of different barrier transparencies, we again make use of Nazarov's boundary conditions \cite{nazarov} which are valid for a non-magnetic but otherwise arbitrary contact. These boundary conditions at $x=\{0,d\}$ may in the present case then be written as\cite{yokoyamaj}
\begin{align}\label{eq:boundary}
L(R_B/R_F)\partial_x\theta &= \pm I_\pm[\mathcal{F}\cos\theta\cos(\psi \mp \phi/2)-\mathcal{G}\sin\theta],\notag\\
L(R_B/R_F)\sin\theta\partial_x\psi &= \mp I_\pm \mathcal{F}\sin(\psi \mp \phi/2),
\end{align}
where the upper (lower) sign is valid at $x=d$ ($x=0$). We have defined the following quantities:
\begin{align}
&I_\pm = \langle4T/[A_\pm(1 + \mathcal{F}^2 + \mathcal{G}^2)]\rangle,\; A_\pm = (2-T) \notag\\
&+2T[\mathcal{F}\sin\theta\cos(\psi\mp\phi/2) + \mathcal{G}\cos\theta]/(1+\mathcal{F}^2+\mathcal{G}^2).
\end{align}
and denoted the phase in the right and left superconductor as $\pm\phi/2$, giving rise to a total phase difference of $\phi$. Above, 
\begin{align}
\mathcal{G} = \varepsilon/\sqrt{\varepsilon^2-\Delta^2},\; \mathcal{F} = \Delta/\sqrt{\Delta^2-\varepsilon^2}
\end{align}
denote components of the bulk Green's function in the superconductor. Eqs. (\ref{eq:usadel}) and (\ref{eq:boundary}) constitute two coupled, second order, nonlinear differential equations with boundary conditions and may be used to numerically solve for $\theta=\theta(x,\varepsilon)$. Once $\theta(x,\varepsilon)$ is obtained, the retarded Green's function is specified everywhere in the diffusive ferromagnet. The current-density may then be calculated by 
\begin{align}
j = -(N_F|e|D/4)\int^{\infty}_{-\infty} \text{d}\varepsilon \text{Tr}\{\hat{\rho}_3 (\check{g}\partial_x \check{g})^\text{K}\},
\end{align}
where $N_F$ is the density of states at Fermi level in the normal state, $|e|$ is the electronic charge, $\hat{\rho}_3$ is a Pauli matrix in particle-hole space, $\check{g}$ is the full Green's function in Keldysh$\otimes$particle-hole$\otimes$spin space, and the superscript 'K' denotes the Keldysh component. The current is obtained by $I=jS_A$, where $S_A$ is the surface area of the junction, and the critical current is defined as $I_c =$ max$\{I(\phi)\}$. The $I_cR_N$-product can now be computed, where $R_N = 2R_B+R_F$. 
In the actual calculations and numerical implementation, we employ the Matsubara representation $\varepsilon\to \i\omega$, and parametrize the quasiclassical Green's functions with the quantity $\Phi_\omega$ as
\begin{align}\label{eq:matgreen}
f_\omega &= g_\omega\Phi_\omega/\omega=\sin\theta\e{-\i\psi}, \; f^*_{-\omega} = \Phi^*_{-\omega}g_\omega/\omega = \sin\theta\e{\i\psi},\notag\\
g_\omega &= \omega/\sqrt{\omega^2 + \Phi_\omega\Phi^*_{-\omega}} = \cos\theta.
\end{align}
We may now express Eqs. (\ref{eq:usadel}) and (\ref{eq:boundary}) in terms of the fermionic frequency $\omega=(2n+1)\pi/\beta$, $n=0,1,2\ldots$, and the Green's functions in Eq. (\ref{eq:matgreen}).
\par
For later purposes, we introduce dimensionless measures of the Thouless energy and inverse spin-flip scattering lifetime: $\mathcal{E}=\varepsilon_T/\Delta = D/(d^2\Delta)$ and $g=\gamma/\Delta$. For simplicitly, we will also neglect the spatial depletion of the superconducting order parameter near the interfaces. 

\section{Results}\label{sec:results}
We now proceed to present our results for the DOS and Josephson current in detail. In the first part of this section, we consider a diffusive ferromagnet/superconductor bilayer, while in the second part we investigate a diffusive superconductor/ferromagnet/superconductor junction.

\subsection{Density of states in ferromagnet/superconductor junction}
We will divide our results for the DOS into subsections to clarify the role of the spin-dependent scattering for different exchange fields and film thicknesses. 
In the spirit of Ref.~\onlinecite{zareyan}, we will consider the cases \textit{i)} $h\lessim\Delta$, \textit{ii)} $h\gtrsim \Delta$, \textit{iii)} $h\gg \Delta$. For $h\gg\Delta$, the proximity effect in the ferromagnet is weak unless the Thouless energy is very high $(\varepsilon_T \gg \Delta)$. Throughout 
the rest of this paper, we fix $\gamma_L = 100$ and $\gamma_R = 1$. This corresponds to a scenario where the normal metal reservoir effectively 
acts as a very strong insulating barrier, mimicking a vacuum boundary. Since the resistance of the N/F interface will constitute the largest contribution to the total resistance, the conductance of the junction will be equivalent to the DOS at the N/F interface in the tunneling limit. We will study an intermediate value $\tau = 0.5$ of the barrier transparency, since this cannot be reached within the usual approximations of either a fully transparent interface $(\tau=1)$ or a tunneling barrier $(\tau \ll 1)$, and contrast this with a strong barrier $\tau=0.1$. These choices for the barrier strength are experimentally the most relevant ones. For each of the cases \textit{i)}-\textit{iii)}, we will investigate two different thicknesses of the ferromagnetic layer, namely $d/\xi = \{0.1,1.0\}$, corresponding to $\varepsilon_T/\Delta = \{100,1\}$. Here, $\xi = \sqrt{D/\Delta}$ is the superconducting coherence length. We comment further 
on our choice of parameters in Sec. \ref{sec:discuss}. For each type of spin-dependent scattering, we define the dimensionless parameter $g$ as a measure of the inverse scattering time. For spin-flip scattering we have $g_\text{so}=0$ and $g=g_\text{sf}/\varepsilon_T$ with $S_z=3, S_{xy}=0$ in Eq. (\ref{eq:usa}) for the uniaxial case while $S_z=S_{xy}=1$ in the isotropic case. For spin-orbit scattering, we have $g_\text{sf}=0$ and $g=g_\text{so}/\varepsilon_T$. Unless otherwise specified, the DOS is calculated right at the interface 
between the normal metal and the ferromagnet, i.e.  $x=0$ (see Fig. \ref{fig:model}). This corresponds precisely to the experimental situation in Ref.~\onlinecite{kontos}.
\par
In what follows, we first present our numerical results for the DOS for the cases \textit{i)}-\textit{iii)} described above. We then investigate and explain the features seen in each of those cases in separate subsections. One of the main conclusions in this section is that the distinction between different types of spin-dependent scattering, \eg spin-flip and spin-orbit scattering, may actually become very important in terms of interpreting the DOS in a ferromagnet/superconductor bilayer. We relate this to the behaviours of the proximity-induced anomalous singlet and triplet Green's functions in Sec. \ref{sec:anomalous}.
\begin{widetext}
\text{ }\\
\begin{figure}[h!]
\centering
\resizebox{1.05\textwidth}{!}{
\includegraphics{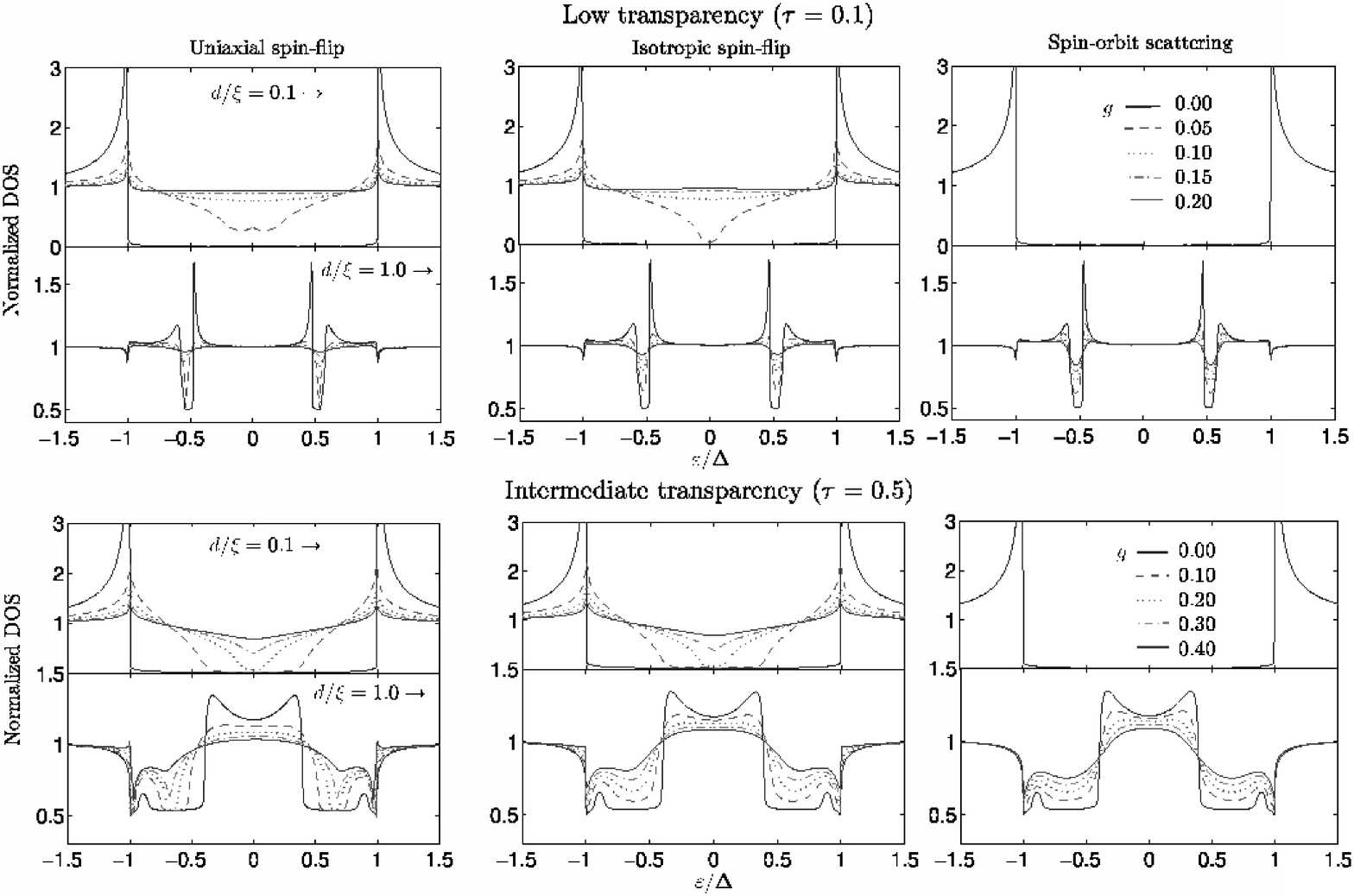}}
\caption{(Color online). The local density of states at $x=0$ (see Fig. \ref{fig:model} for $h/\Delta=0.5$. For each thickness of the ferromagnetic layer and barrier transparency, we study the role of different types of spin-dependent scattering. }
\label{fig:Case1}
\end{figure}
\begin{figure}[h!]
\centering
\resizebox{1.05\textwidth}{!}{
\includegraphics{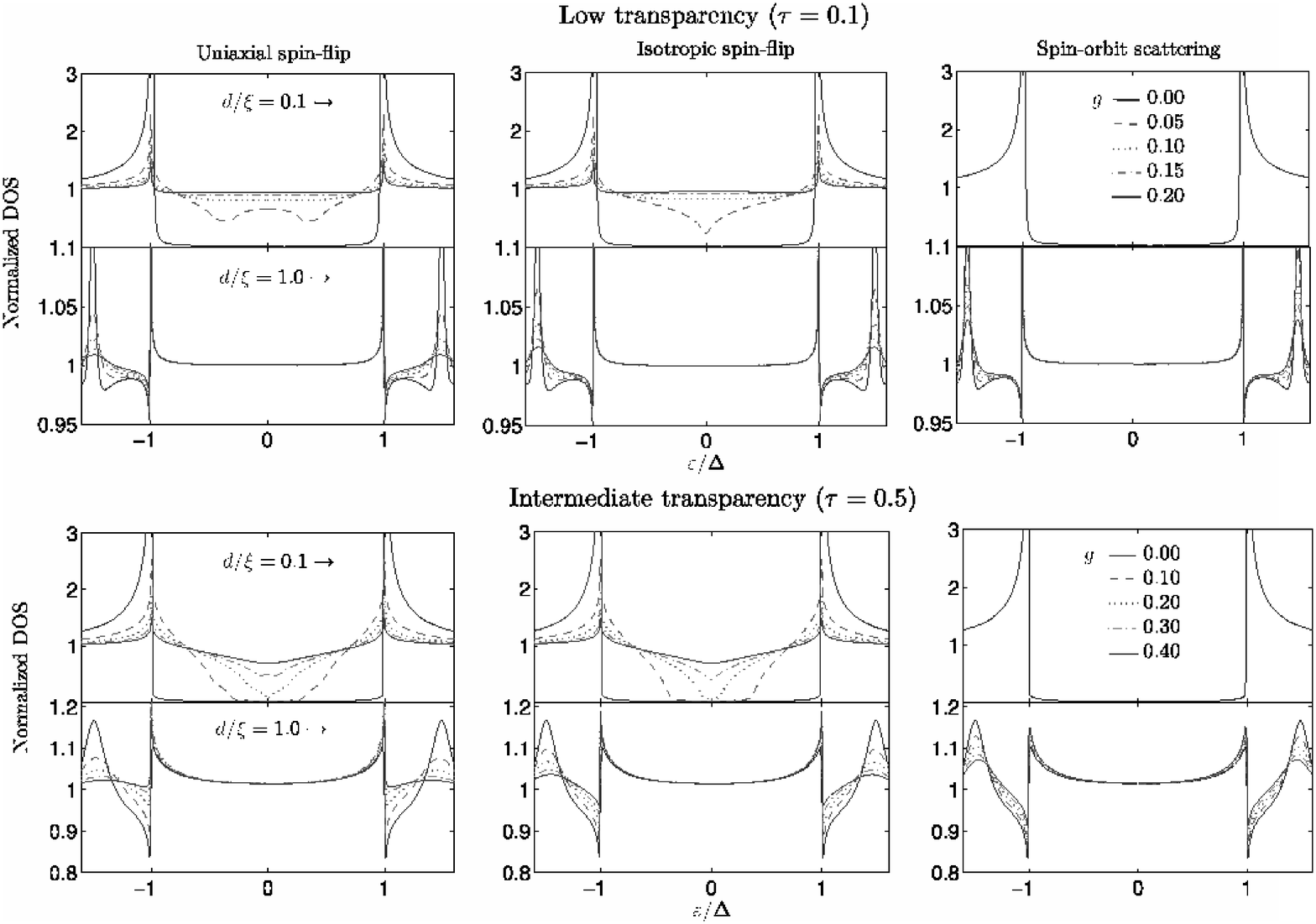}}
\caption{(Color online). The local density of states at $x=0$ (see Fig. \ref{fig:model} for $h/\Delta=1.5$. For each thickness of the ferromagnetic layer and barrier transparency, we study the role of different types of spin-dependent scattering. }
\label{fig:Case2}
\end{figure}
\end{widetext}

\subsubsection{Case \textit{i)}: $h\lessim\Delta$}
We first consider the case of a very weak exchange field, $h/\Delta = 0.5$, which splits the DOS for spin-$\uparrow$ and 
spin-$\downarrow$ electrons. A peculiar feature that may arise in a ferromagnet/superconductor junction is an enhancement of 
the DOS at zero energy, manifested as a peak. This issue was addressed in Ref.~\onlinecite{zareyan,yokoyama,yokoyama07}, and has also been 
experimentally observed in Ref.~\onlinecite{kontos}. 
As seen in Fig. \ref{fig:Case1}, the ferromagnet becomes fully proximized by the superconductor in the thin-layer case $d/\xi=0.1$, and the DOS is equivalent to the bulk of the superconductor (in the absence of spin-dependent scattering). In fact, the DOS is virtually unaltered compared 
to the paramagnetic case for the short-junction regime regardless of whether we consider the tunneling or intermediate transparency case. However, an important distinction between different types of spin-dependent scattering becomes evident in the thin-layer case $d/\xi=0.1$. The effect of spin-flip scattering, whether it is uniaxial or isotropic, is that the DOS-gap closes upon increasing $g$. However, increasing the spin-orbit scattering rate does not affect the DOS in any way. We investigate this issue in more detail in Sec. \ref{sec:minigap}.
\par
The situation changes dramatically when going to a wider junction regime, here modelled by $d/\xi=1.0$. In the tunneling limit, a minigap-like structure opens up in the DOS at energies $\varepsilon=\pm h$. This can be understood as a result of the exchange-splitting in the ferromagnet which shifts the energies, and hence the density of states, for majority and minority spins by $\pm h$. Increasing the barrier transparency to $\tau=0.5$, the minigap-like structure is retained and widened, which is reasonable since the proximity effect becomes larger and the minigap scales with $\tau$. Another feature that is seen in the $d/\xi=1.0$ case is the appearance of a zero-energy peak (ZEP) in the DOS. The peak protrudes with increasing spin-flip scattering and is actually split into two in the absence of magnetic impurities. 
In Ref.~\cite{gosu2}, where the tunneling limit was applied to the F/S interface, no zero-energy peak was found to appear in the spectra. In the clean limit, however, a zero-energy peak was found to appear in the DOS for a wide range of parameters \cite{zareyan}. Although the appearance of a ZEP has been investigated in previous work, none of these considered the effect of spin-dependent scattering. In order to understand the ZEP feature seen in Fig. \ref{fig:Case1}, we study the dependence of the anomalous Green's functions on spin-selective scattering in Sec. \ref{sec:anomalous}.

\subsubsection{Case \textit{ii)}: $h\gtrsim \Delta$}
Next, we consider the case $h/\Delta = 1.5$, shown in Fig. \ref{fig:Case2}. As in the previous case, the DOS is almost unaffected by 
the presence of an exchange field in the short-junction regime $d/\xi=0.1$. Also, the gap in the DOS shows a remarkable resiliance towards increasing the spin-orbit scattering. However, for $d/\xi=1.0$ the DOS is strongly modified and displays two peaks located at $\varepsilon=\pm\Delta$ and $\varepsilon=\pm h$, 
respectively. Increasing the concentration of magnetic impurities suppresses these peaks. In the clean limit, 
Zareyan \etal found a similar development of the DOS with increasing Thouless energies (see Fig. 4 in Ref.~\onlinecite{zareyan}). Note 
that the suppression of the proximity-induced features in the DOS due to the superconductor is now stronger for a given $d/\xi$ as 
compared to the case $h<\Delta$ (Fig \ref{fig:Case1}). In general, increasing the value of $h$ leads to a smaller magnitude of the proximity-induced anomalous Green's functions (see also Fig. \ref{fig:anomalous}), which in turn causes the normalized DOS in the ferromagnetic layer to deviate less from unity.

\subsubsection{Case \textit{iii)}: $h\gg \Delta$}
Finally, we investigate the case $h \gg \Delta$. If the ferromagnetic layer is an alloy of the type Cu$_{1-x}$Ni$_x$, a reasonable value 
of the exchange field may be found in the range $h=10-50$ meV. Here, we consider $h/\Delta=15$ and $h/\Delta=50$. With increasing value of the exchange field, the proximity effect becomes weaker. Therefore, we restrict our attention to the short-junction regime $(d/\xi=0.1$) with an intermediate barrier-transparency $(\tau=0.5)$ since the DOS deviates little from unity in the tunneling regime and for wide ferromagnetic layers. 
\par
For $h/\Delta=15$, the usual peak at $\varepsilon=\Delta$ is present, and a minigap structure is seen in the absence of spin-flip scattering. Interestingly, increasing spin-flip scattering not only closes the minigap, but actually causes the DOS to develop a peak at zero energy. This may be understood by considering a subtle interplay between the singlet and triplet components of the proximity-induced anomalous Green's function in the ferromagnetic layer, and we delay a detailed explanation to Sec. \ref{sec:anomalous}.
\par
For $h/\Delta=50$, all peaks and minigap features are now absent for $|\varepsilon|<\Delta$, and the only feature remaining in the spectrum is a dip at $\varepsilon=\Delta$. In this case, the qualitative effect of the different types of spin-dependent scattering is the same. Upon increasing $d/\xi$ even further $(d/\xi\gg 1)$, corresponding to a weaker proximity effect, one finds that the correction to the DOS oscillates upon increasing the spin-flip scattering 
rate, in contrast to the monotonous decay that might have been expected \cite{linder07}. These oscillations are a result of the modified oscillation length of the proximity-induced anomalous Green's function in the F region \cite{bergeretRMP,buzdin}.
\par
Also note that although the corrections to the normal-state DOS diminish 
rapidly for $h\gg\Delta$ upon increasing $d/\xi$, the combination of lock-in detection with an ultra-low noise DC/AC mixer permits the 
resolution of structures in the DOS up to a factor $10^{-4}$ smaller than the background conductance \cite{kontos}.
\begin{widetext}
\text{ }
\begin{figure}[h!]
\centering
\resizebox{0.7\textwidth}{!}{
\includegraphics{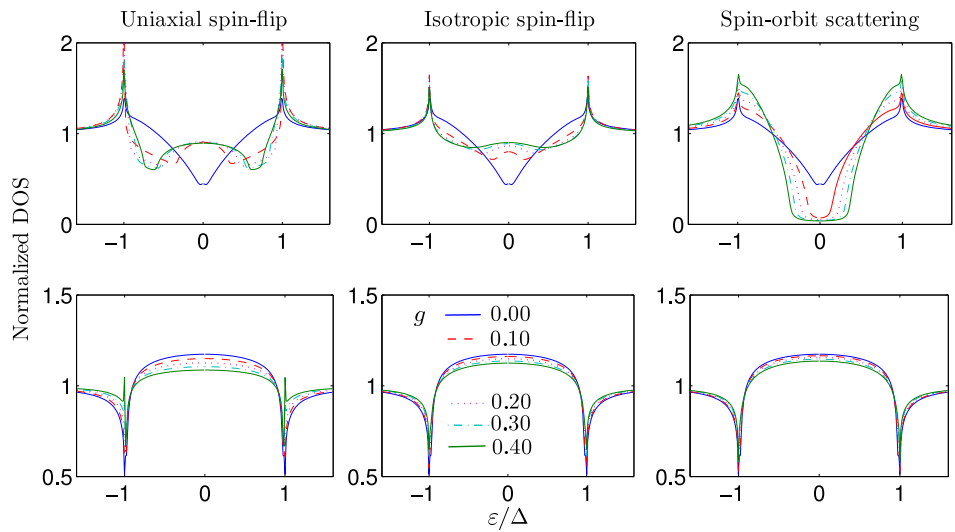}}
\caption{(Color online). Local density of states at $x=0$ (see Fig. \ref{fig:model}) for $h/\Delta=15$ (upper row) and $h/\Delta=50$ (lower row). Here, we have fixed $d/\xi=0.1$ and $\tau=0.5$ and study the role of different types of spin-dependent scattering.}
\label{fig:Case3}
\end{figure}
\end{widetext}

\subsubsection{Zero-energy peak}\label{sec:zep}

\begin{figure}[h!]
\centering
\resizebox{0.48\textwidth}{!}{
\includegraphics{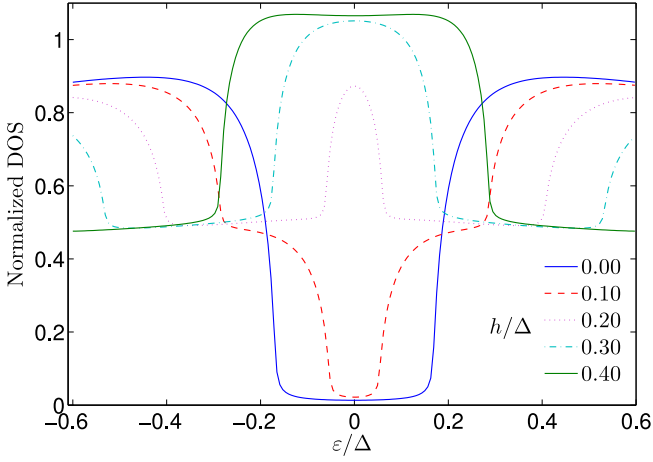}}
\caption{(Color online). Plot of the DOS of a ferromagnet/superconductor junction with increasing exchange field $h$, using $d/\xi=1.0$ and $\tau=0.5$. Here, we have set the spin-dependent scattering to zero.}
\label{fig:ZEP}
\end{figure}

It is instructive to consider the development of the zero-energy peak in the DOS. In Fig. \ref{fig:ZEP}, we plot the DOS around zero 
energy for $d/\xi=1.0$ for increasing values of the exchange field $h$. The gradual formation of a zero-energy peak is clearly observed, 
and the peak flattens out when $h$ becomes sufficiently large. The physical reason for this is intriguing. Yokoyama \etal \cite{yokoyama07} 
have related this phenomenon directly to the proximity-induced odd-frequency pairing component of the anomalous Green's function in the 
ferromagnet (see also Refs. \cite{Braude,Asano}). In these works, it is shown that the DOS in the ferromagnet is enhanced due to the emergence of the proximity-induced odd-frequency pairing. Since we are considering the isotropic part (with respect to momentum) of the Green's function due to the angular averaging 
in the dirty limit, one would perhaps naively expect that only the singlet component should be present. This is because the singlet 
anomalous Green's function is usually taken to be even under inversion of momentum, while the triplet components are taken to be odd 
under inversion of momentum. However, another possibility exists that allows for the presence of triplet correlations in the ferromagnet, 
involving a change of sign of the superconducting order parameter under inversion of frequency. This type of pairing has been dubbed \textit{odd-frequency pairing} in the literature.\cite{bergeretRMP} Inversion of frequency is equivalent to an exchange of 
(relative) time coordinates for the field operators, since $\varepsilon$ is the Fourier transform of the relative time coordinate 
$t\equiv t_1-t_2$. Note that although even-frequency triplet correlations are destroyed in the dirty limit due to the isotropization 
stemming from impurity scattering, odd-frequency triplet correlations may persist, since these do not vanish under angular averaging. 
\par
It remains to be clarified how the ZEP in a ferromagnet/superconductor junction is affected by spin-dependent scattering. To investigate this, we plot the DOS in Fig. \ref{fig:zep_spin} for a fixed exchange field $h/\Delta=0.3$ which gives a ZEP in the absence of spin-dependent scattering, and then successively increase the scattering rate. It is seen that the effect of increasing the spin-flip scattering rate (both uniaxial and isotropic) is a suppression of the proximity-induced features in the DOS. Qualitatively, the same occurs upon increasing the spin-orbit scattering rate (right panel of Fig. \ref{fig:zep_spin}), but an interesting difference is that the peak is eventually transformed into a dip at $\varepsilon=0$. Increasing the spin-orbit scattering rate further $(g\gg1)$ leads to a fully developed minigap in the DOS. We propose an explanation of this peculiar phenomenon in the following section.
\begin{widetext}
\text{ }\\
\begin{figure}[h!]
\centering
\resizebox{0.75\textwidth}{!}{
\includegraphics{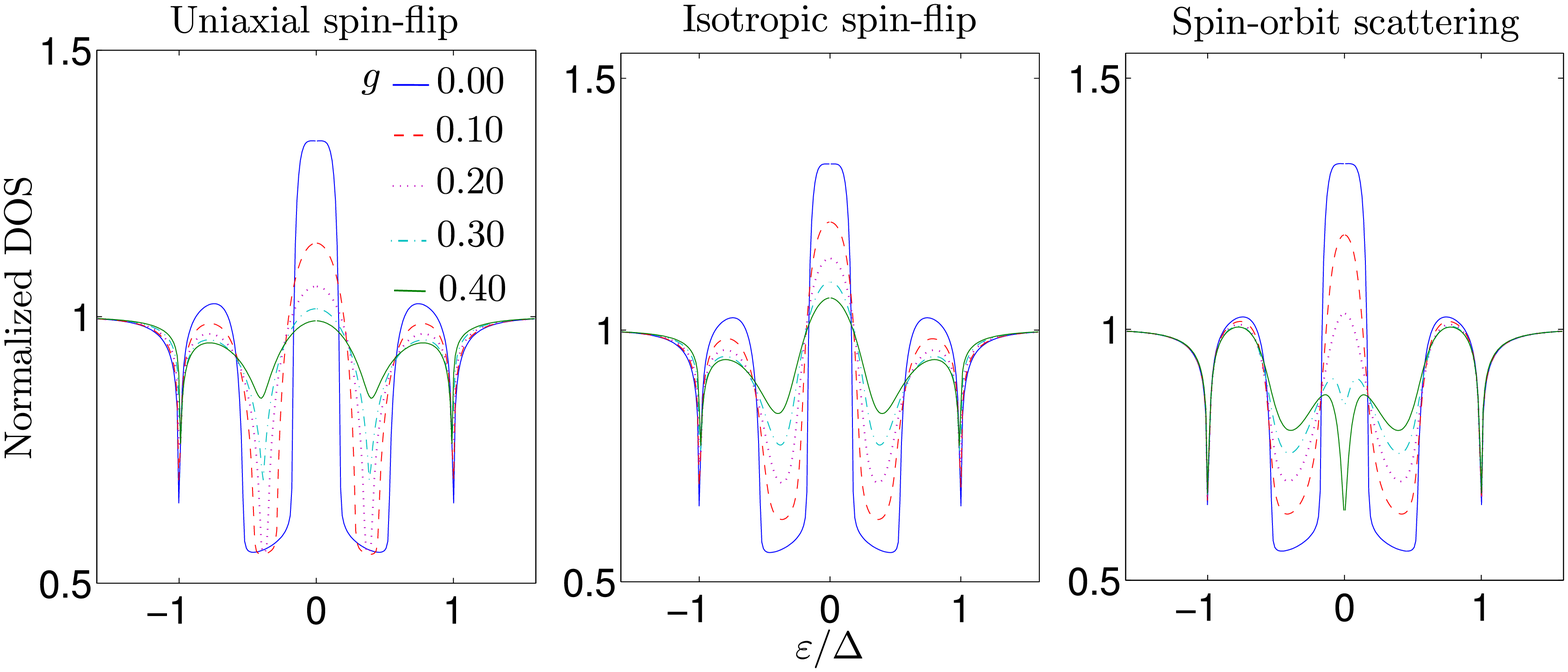}}
\caption{(Color online). Plot of the DOS of a ferromagnet/superconductor junction for a fixed exchange field $h/\Delta=0.3$ which gives a ZEP in the absence of spin-dependent scattering. We fix $d/\xi=1.0$ and $\tau=0.5$. }
\label{fig:zep_spin}
\end{figure}
\end{widetext}

\subsubsection{Anomalous Green's functions}\label{sec:anomalous}
In order to understand the interplay between the singlet and triplet components of the induced superconducting anomalous Green's function in the F region with regard to the zero-energy behaviour of the DOS, consider the anomalous Green's functions defined as:
\begin{align}\label{eq:defano}
f_s(\varepsilon,x) &= [\sinh\theta_\uparrow(\varepsilon,x) - \sinh\theta_\downarrow(\varepsilon,x)]/2, \notag\\
f_t(\varepsilon,x) &= [\sinh\theta_\uparrow(\varepsilon,x) + \sinh\theta_\downarrow(\varepsilon,x)]/2.
\end{align}
At zero energy $\varepsilon=0$, one finds that Re$\{f_s\}$=Im$\{f_t\}$=0. The reader is reminded of the relation between the anomalous Green's functions and the DOS, which is a physical observable: the total DOS is given as $N = \sum_\sigma \text{Re}\{\cosh\theta_\sigma\}/2$. Also note that the singlet and triplet components differ not only in their spin-symmetry, but also with respect to their energy-dependence (even and odd, respectively), as noted in Sec. \ref{sec:zep}. 
\par
\begin{figure}[h!]
\centering
\resizebox{0.48\textwidth}{!}{
\includegraphics{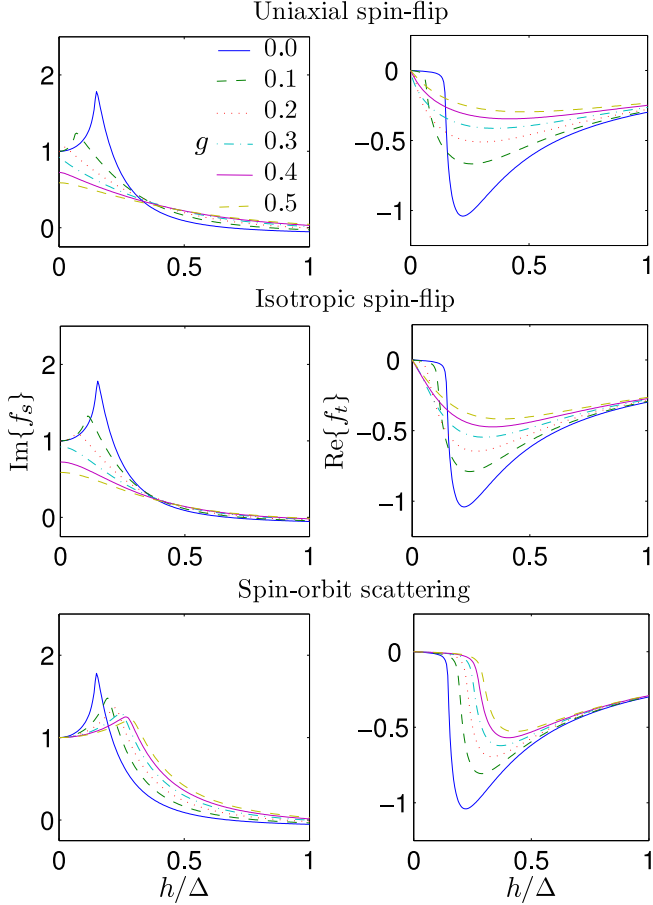}}
\caption{(Color online). Plot of the singlet (Im$\{f_s\}$) and triplet (Re$\{f_t\}$) anomalous Green's functions induced in the ferromagnet for $\varepsilon=0$ right at the N/F interface $(x=0)$. We have used $\tau=0.5$ and $d/\xi=1.0$.}
\label{fig:anomalous}
\end{figure}

In Fig. \ref{fig:anomalous}, we plot Im$\{f_s\}$ and Re$\{f_t\}$ as measures of the singlet and triplet induced proximity Green's function in the ferromagnet right at the N/F interface $(x=0)$. We consider the effect of each type of spin-dependent scattering separately. As seen, the peak in the singlet component occuring at a finite value of $h$ vanishes upon increasing spin-flip scattering rate, and the decay eventually becomes monotonuous for $g \grtsim 0.3$. However, this is not the case for the triplet component: the maximum value of $|$Re$\{f_t\}|$ occurs at a finite value of $h$ even upon increasing the spin-dependent scattering rate. Therefore, the triplet component may become similar in magnitude to the singlet component even for weak exchange fields if the spin-flip scattering rate is sufficiently large. This explains the appearance of a zero-energy peak in the plot for $d/\xi=1.0$ in Fig. \ref{fig:Case2} upon increasing $g$.
\par
Another interesting feature can be observed in Fig. \ref{fig:anomalous}. From the plots, it is seen that the anomalous Green's functions (both singlet and triplet) for a given exchange field may actually become larger upon increasing the rate of spin-flip scattering. This becomes evident around $h/\Delta=0.1$ when comparing the lines for $g=0.0$ with $g=0.1$. Thus, we have at hand the opportunity to see an \textit{enhanced proximity-effect by spin-flip scattering} due to the increased magnitude of $|f_{s,t}|$ at low values of $h$. The enhanced proximity-effect actually becomes very pronounced also in the case of bulk odd-frequency superconductors \cite{future}.
\par
From Fig. \ref{fig:anomalous}, we may actually also find an explanation for the remarkable behaviour of the DOS upon increasing the spin-orbit scattering rate in Fig. \ref{fig:zep_spin}. For concreteness, let us first focus on the regime $h/\Delta \simeq 0.3$ corresponding to Fig. \ref{fig:zep_spin}. Let us compare the plots for spin-flip and spin-orbit scattering. For both uniaxial and isotropic spin-flip scattering, it is seen that the singlet component does not change much in magnitude, while the triplet component is suppressed. However, the singlet and triplet components are still comparable in magnitude for $g=0.5$, which accounts for the suppression of the minigap feature which is due to the singlet component. The situation is markedly different for the spin-orbit scattering. Now, the singlet component is actually enhanced in magnitude (for $h/\Delta \simeq 0.3)$ while the triplet component is reduced very strongly. In fact, the triplet component becomes essentially zero for very large spin-orbit scattering rates, while the singlet component still may have a considerable magnitude. This explains the appearance of a minigap in Fig. \ref{fig:zep_spin} upon increasing the spin-orbit scattering rate: the singlet component, which is responsible for the minigap, increases while the triplet component decreases. 
\par
\begin{figure}[h!]
\centering
\resizebox{0.48\textwidth}{!}{
\includegraphics{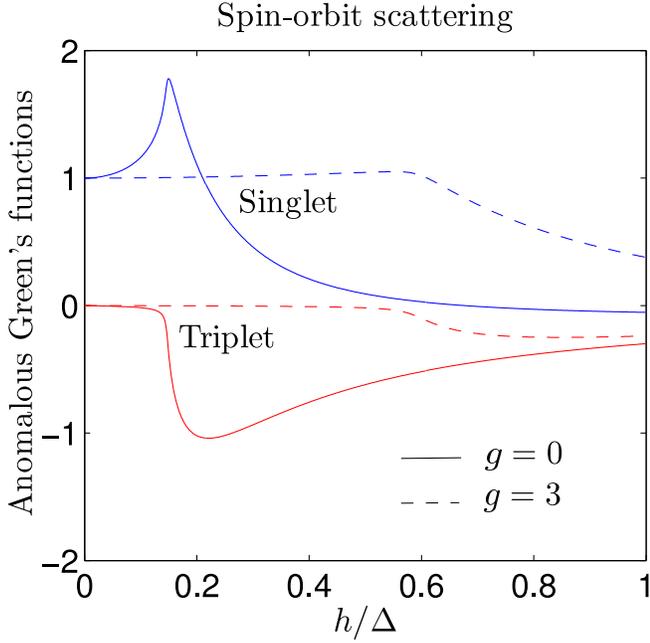}}
\caption{(Color online). Plot of the singlet (Im$\{f_s\}$) and triplet (Re$\{f_t\}$) anomalous Green's functions induced in the ferromagnet for $\varepsilon=0$ right at the N/F interface $(x=0)$. We have used $\tau=0.5$ and $d/\xi=1.0$.}
\label{fig:singletvstriplet}
\end{figure}

The above discussion was restricted to $h/\Delta \simeq 0.3$ for concreteness, but the results nevertheless allude to a much more general principle: the spin-orbit scattering is much more detrimental for the triplet component than the singlet component. To elucidate this feature, consider Fig. \ref{fig:singletvstriplet}, where we compare the proximity-induced singlet and triplet component in the ferromagnetic layer. We set $d/\xi=1.0$, since for $d/\xi\ll 1$ the singlet component completely dominates the triplet component (see Fig. \ref{fig:Case1} and \ref{fig:Case2}). Also, we set $\tau=0.5$, but we checked that the qualitative features are identical for lower barrier transparencies. As seen from the figure, the singlet component is much more robust towards increased spin-orbit scattering than the triplet component. The latter only becomes appreciable in magnitude when the exchange field becomes of the same order or larger than $\Delta$ when the spin-orbit scattering range is large, $g\gg1$. Note that in contrast, both the singlet and triplet components are strongly reduced with increasing spin-flip scattering. 
\par
In the absence of an exchange field, it is clear from Fig. \ref{fig:anomalous} that the singlet component is completely independent of the rate of spin-orbit scattering. Although we have only shown this explicitly for $\varepsilon=0$, we have numerically confirmed that the magnitude of the singlet component at $h=0$ remains unchanged upon increasing $g$. On the other hand, the triplet component is always zero at $h=0$, but remains very close to zero upon increasing the exchange field $h$ for high values of $g$. A natural question arises: why is the singlet component insensitive to spin-orbit scattering while the triplet component depends strongly on it? 
\par
The answer to this question is illuminated by considering the Usadel equations in the ferromagnetic region in limiting cases. From Eq. (\ref{eq:usa}), we obtain several important properties:
\begin{itemize}
\item In the presence of uniaxial spin-flip scattering ($S_z=3, S_{xy}=0$), by taking the limit of $\tau_\text{sf} \to 0$ we get $\theta_\uparrow=\theta_\downarrow=0$. Then both singlet and triplet components are suppressed. This is also the case for isotropic spin-flip scattering ($S_z=S_{xy}=1$). 
\item With in-plane spin-flip scattering ($S_z=0$), we get $\theta_\uparrow=\theta_\downarrow$ when taking the limit $\tau_\text{sf} \to 0$. Then, the singlet component should vanish as seen from definition in Eq. (\ref{eq:defano}). 
\item In the presence of pure spin-orbit scattering ($S_z= S_{xy}=0$), by taking limit of $\tau_\text{so} \to 0$ we get $\theta_\uparrow=-\theta_\downarrow$. Then, the triplet component should vanish as seen from the definition in Eq. (\ref{eq:defano}). 
\end{itemize}
Therefore, we find that uniaxial and isotropic spin-flip scattering is harmful to both singlet and triplet componets while in-plane spin-flip scattering and spin-orbit scattering are detrimental to the singlet and triplet components, respectively. 
\par
These properties are also obtained by using linearized Usadel equations in the ferromagnetic region, which may be formally obtained from Eq. (\ref{eq:usa}) by letting $\theta_\sigma \to f_\sigma$ with $\sigma=\pm$, and assuming that $|f_\pm|\ll1$: 
\begin{align}
\partial_x^2 f_t \pm \partial_x^2 f_s + A_\pm f_t \pm B_\pm f_s = 0,
\end{align}
where we have introduced
\begin{align}\label{eq:AB}
A_\pm &= \frac{1}{D}\Big[2\i(\varepsilon\pm h) - g_\text{so} - \frac{g_\text{sf}S_z}{2}\Big],\notag\\
B_\pm &= \frac{1}{D}\Big[2\i(\varepsilon\pm h) - \frac{g_\text{sf}(2S_{xy}+S_z)}{2}\Big].
\end{align}
As seen, the spin-orbit scattering rate only enters in the coefficient associated with the triplet component. Notice that in absence of an exchange field $h$, which renders $f_t=0$, the linearized Usadel equation is independent of the spin-orbit scattering rate. On a microscopical level, it is clear that the different dependence on spin-orbit scattering for the singlet and triplet anomalous Green's functions originates from the fundamental symmetries of these wavefunctions. It should be emphasized that \textit{both} the singlet and triplet components are strongly affected by magnetic impurities, i.e. spin-flip scattering, as seen from Eq. (\ref{eq:AB}).
\par
The above analysis emphasizes the importance of distinguishing between different types of spin-dependent scattering in terms of understanding the behaviour of the DOS in a ferromagnet/superconductor bilayer. In particular, we have shown that the effect of spin-orbit scattering may differ fundamentally from spin-flip scattering (originating \eg from magnetic impurities), and that this is manifested in the interplay between the singlet and triplet anomalous Green's function in the ferromagnetic layer. 

\subsubsection{Minigap and spin-orbit scattering}\label{sec:minigap}
Finally, we consider the effect of spin-dependent scattering on the interesting manifestation of the proximity-induced superconducting correlations in a normal/superconductor junction: the minigap \cite{mcmillan}. 
We first briefly recapitulate the results in the paramagnetic case. When a normal metal is placed in close proximity to a superconductor, a minigap $\Delta_g$ opens up in the DOS of the normal metal. This minigap roughly scales like $\Delta_g \sim \varepsilon_T\tau/\gamma_R$, and was originally studied by McMillan in a tunneling model of a normal/superconductor junction \cite{mcmillan}. The minigap is 
defined as an almost complete suppression of the quasiparticle DOS in a given energy interval. 
The minigap is a consequence of the 
effective backscattering that quasiparticles incident on the superconducting interface from the normal side experience due to 
the presence of impurities. Consequently, the probability for transmission increases such that the DOS is nonzero for 
$\Delta_g <\varepsilon <\Delta$. For nearly perfect transparency of the interface, $\Delta_g$ can become close to $\Delta$ in 
magnitude. For a tunneling barrier $\tau\ll 1$, the minigap becomes very small. 
\par
An interesting issue is how the minigap is affected by spin-dependent scattering. We provide numerical results to elucidate this question in Fig. \ref{fig:minigap}. As seen from the plots, the influence of uniaxial spin-flip scattering is a gradual suppression of the proximity-induced features in the DOS,  
consistent with previous results \cite{belzig, golubov1988,Volkov,Yip,Yoko}. The effect of isotropic spin-flip scattering is virtually identical to the uniaxial case. However, quite surprisingly, the minigap shows strong resiliance towards increasing spin-orbit scattering as seen in Fig. \ref{fig:minigap}. In fact, we find that even upon increasing $g$ to values $\gg\Delta$, the minigap persists in the DOS as long as the exchange field is absent. Upon increasing the exchange field to values $h\gg\Delta$, the minigap slowly begins to close when the spin-orbit scattering rate becomes large. This feature may be understood by again resorting to Fig. \ref{fig:anomalous} and Fig. \ref{fig:singletvstriplet}, where a detailed study of the singlet and triplet components of the anomalous Green's function was conducted. In general, the spin-orbit scattering was shown to be detrimental for the triplet component while the singlet component still remained at a considerable magnitude even for $g\gg1$. However, the triplet component was strengthened upon increasing the exchange field $h$, which explains why the minigap would close for larger values of $h$ upon increasing the scattering rate. This behaviour should be contrasted with spin-flip scattering, whether it is isotropic or uniaxial, which invariably suppresses both the singlet and triplet components of the proximity-induced anomalous Green's function.
\begin{widetext}
\text{ }\\
\begin{figure}[h!]
\centering
\resizebox{0.75\textwidth}{!}{
\includegraphics{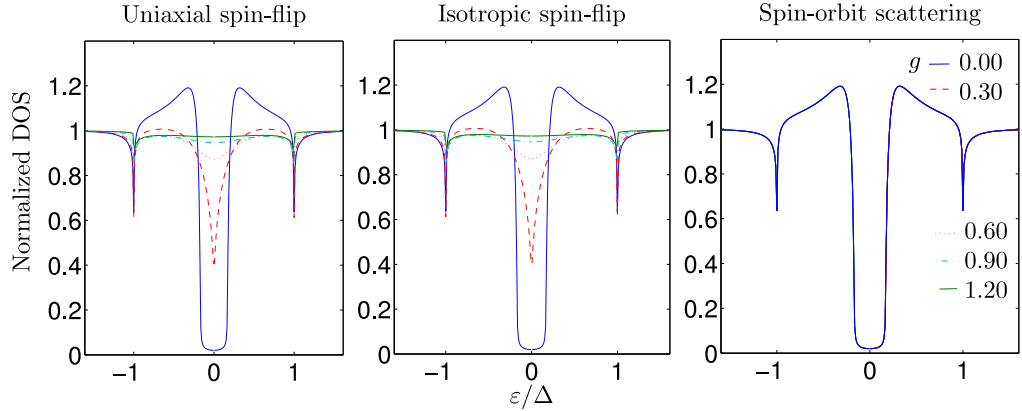}}
\caption{(Color online). Plot of the DOS at $x=0$ for a normal metal/superconductor junction using $d/\xi=1.0$ and $\tau=0.5$. As seen, spin-flip scattering closes the minigap, whether it is uniaxial or isotropic. However, the minigap shows strong resiliance against increasing spin-orbit scattering.}
\label{fig:minigap}
\end{figure}
\end{widetext}

\subsection{Josephson current}
Having concluded our study of the DOS in the F/S bilayer, we now turn to the supercurrent in a S/F/S junction. 
In the following, we fix $R_B/R_F = 5$ where $R_B$ is the resistance of the barrier at the interfaces and $R_F$ is the resistance of the diffusive ferromagnetic layer. Also, we consider an exchange field $h/\Delta=10$, which should correspond to a typical weak ferromagnetic alloy such as Cu$_{1-x}$Ni$_x$. In the following, we are particularly interested in examining the possibility of obtaining a strong deviation from the usual sinusoidal current-phase relationship. This may lead to the opportunity of creating a supercurrent-switching device, which we elaborate on below. We remind the reader that all the quantities used below were introduced and defined in Sec. \ref{sec:theory}.
\par
\begin{figure}[h!]
\centering
\resizebox{0.40\textwidth}{!}{
\includegraphics{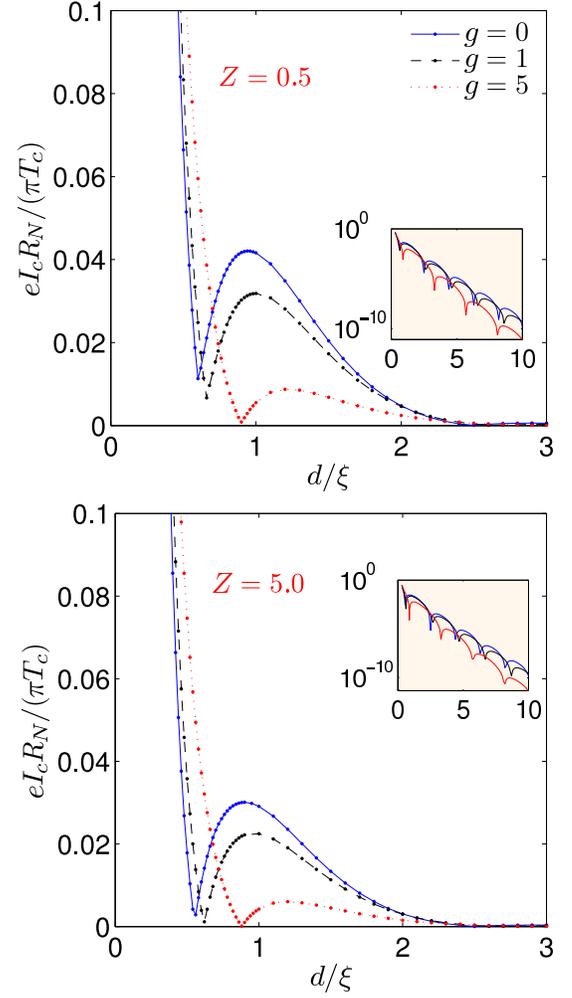}}
\caption{(Color online) Plot of junction-width dependence of the $I_cR_N$-product for an S/F/S Josephson junction in the intermediate $(Z=0.5)$ and low $(Z=5.0)$ barrier transparency regime. The temperature is fixed at $T/T_c=0.1$. For each case, the effect of increasing uniaxial spin-flip scattering is seen to severly reduce the residual value of the critical current at the 0-$\pi$ transition points.}
\label{fig:length}
\end{figure}

Consider first dependence of the $I_cR_N$-product on the normalized junction width $d/\xi$, where $\xi=\sqrt{D/2\pi T_c}$ is the superconducting coherence length. We contrast the results obtained with a barrier of intermediate transparency $(Z=0.5)$ and low transparency $(Z=5.0)$ as the experimentally most relevant cases, shown in Fig. \ref{fig:length}. As seen, the intermediate transparency allows for a finite residual value of the supercurrent at the 0-$\pi$ transition point, where the first harmonic of the current-phase relationship vanishes. The effect of uniaxial spin-flip scattering is seen to be a reduction of the residual value of the supercurrent. Even for intermediate transparency of the barrier, the residual current is reduced to immeasurable values for $g\sim5$, where $g=\gamma/\Delta$ is a measure of the uniaxial spin-flip scattering. This issue has not been addressed previously in the literature, and is useful to complement previous qualitative predictions with a more realistic quantitative analysis.
\begin{figure}[h!]
\centering
\resizebox{0.37\textwidth}{!}{
\includegraphics{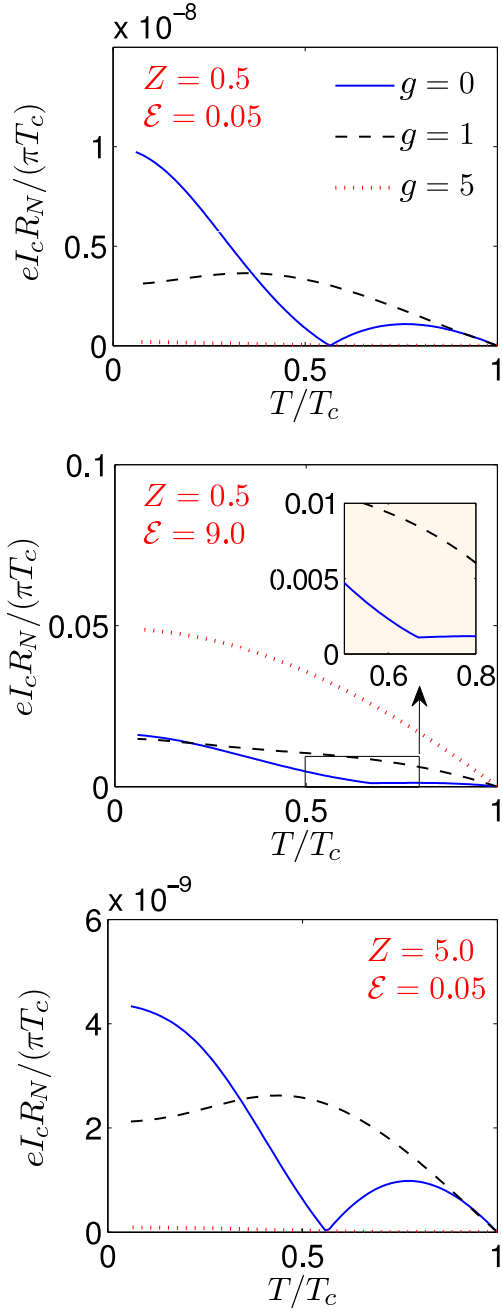}}
\caption{(Color online) Plot of the temperature-dependence of the $I_cR_N$-product for an S/F/S Josephson junction in the intermediate $(Z=0.5)$ and low $(Z=5.0)$ barrier transparency regime for different values of the normalized Thouless energy $\mathcal{E}$. }
\label{fig:temp}
\end{figure}
Also, it is seen from Fig. \ref{fig:length} that the transition points are translated towards higher junction widths upon increasing $g$.
\par
\begin{figure}[h!]
\centering
\resizebox{0.40\textwidth}{!}{
\includegraphics{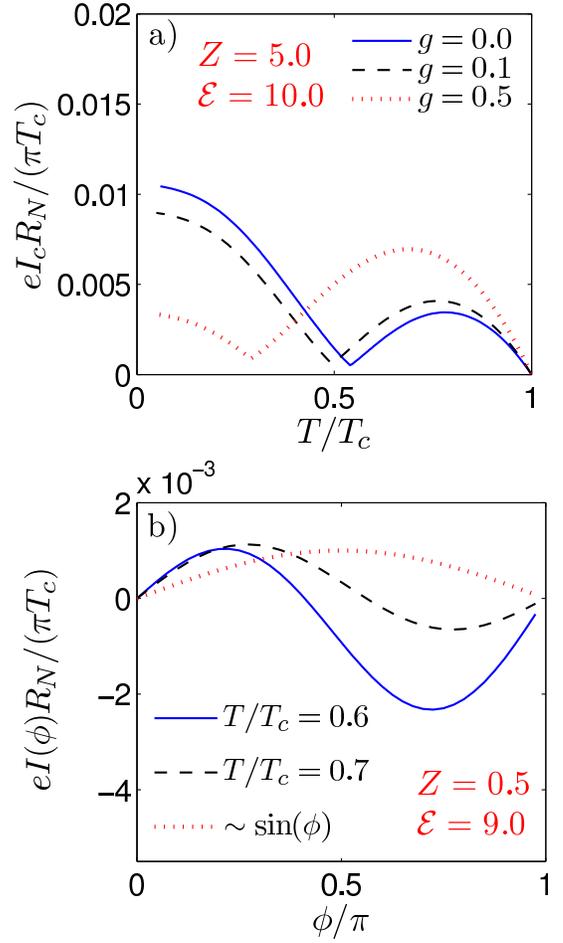}}
\caption{(Color online) a) Plot of the temperature-dependence of the $I_cR_N$-product for weakly increasing spin-flip scattering. b) Plot of the current-phase relationship for the $I_cR_N$-product for an S/F/S Josephson junction, illustrating the strong deviation from the sinusoidal dependence and the 0-$\pi$ transition.}
\label{fig:phase}
\end{figure}

We next consider the temperature-dependence of the $I_cR_N$-product in Fig. \ref{fig:temp}. In the wide-junction regime $\mathcal{E}=\varepsilon_T/\Delta=0.05$, the magnitude of $I_cR_N$ is quite small and a 0-$\pi$ transition occurs for zero spin-flip scattering at $T/T_c\approx 0.55$. However, increasing the value of $g$ is seen to vanish the 0-$\pi$ transition completely. This behaviour is very distinct from the $d/\xi$-dependence shown in Fig. \ref{fig:length}, where increasing spin-flip scattering reduces the residual value of the supercurrent at the transition point, but does not remove the transition completely. Consider now a shorter junction here modelled by $\mathcal{E}=9.0$, corresponding to $d/\xi\approx0.6$, for an intermediate barrier transparency $Z=0.5$. Close examination of the transition point where the current changes sign reveals a small but finite residual value of the supercurrent, in this case for $T/T_c \approx 0.65$. At first sight, the effect of increasing $g$ then appears to amount to a complete removal of the 0-$\pi$ transition point, rather than a suppression of the residual value as in Fig. \ref{fig:length}. However, by increasing $g$ in smaller steps as shown in Fig. \ref{fig:phase}a) with $\mathcal{E}=10.0$, it is seen that the 0-$\pi$ transition \textit{gradually} vanishes. This means that the transition point when considering the temperature-dependence is much more sensitive to spin-flip scattering than the width-dependence of Fig. \ref{fig:length}. It should be noted that the 0-$\pi$ oscillations vanish upon increasing $g$ for the particular choice of the width $d$ (corresponding to a certain $\mathcal{E}$) used here. For another choice of $d$, one might expect to have 0-$\pi$ introduced upon increasing $g$. The main point is nevertheless that the temperature-dependence of the critical current oscillations appear to be more sensitive to spin-flip scattering than the width-dependence of the same oscillations.
\par
The origin of a residual value of the current at the sign-reversal point is a deviation from a purely sinusoidal current-phase relationship. To illustrate the strong deviation from a pure sinusoidal phase-dependence for the $\mathcal{E}=9.0$ case, consider Fig. \ref{fig:phase}b). The 0-$\pi$ transition can clearly be discerned from the plot. Right before the transition ($T/T_c=0.6$), the maximum value of the current occurs at a negative value for $I_cR_N$. After the transition ($T/T_c=0.7$), the maximum value occurs for a positive value of $I_cR_N$. For comparison, we have plotted a pure sinusoidal phase-dependence. Fig. \ref{fig:phase}b) shows that that higher harmonics [in this case $\sin(2\phi)$] dominate near the sign reversal point, giving rise to the residual value of the current.

\section{Discussion}\label{sec:discuss}
In our calculations, we have neglected the depletion of the superconducting and magnetic order parameter in the vicinity of the F/S 
interface. In the low transparency case, this is permissable \cite{bruder}. In the high transparency case, the depletion may be substantial. However, 
the qualitative features obtained in the DOS are known to be the same even if this depletion is taken into account - the characteristic 
features are often simply shifted in energy from the bulk value $\varepsilon=\Delta$ to a reduced value $\tilde{\Delta} < \Delta$. Our approximation consisting of using the bulk Green's function in the superconducting regions is well justified also if we assume that the superconducting region is much less disordered than the ferromagnetic layer.
Also, 
we have employed an effectively one-dimensional model to account for the proximity-effect. In the present system, we are concerned with isotropic order parameters ($s$-wave superconductivity), such that performing the same calculations in 2 or 3 dimensional will yield similar results since the averaging over angles will have virtually no effect.
\par
Let us also comment on our choices of Thouless energy in the ferromagnetic region. The actual thicknesses employed in experiments on both 
F/S bilayers and S/F/S Josephson junctions vary greatly, but may in some cases be as small as $<$ 20 nm \cite{kontos, ryazanov}. Now, the 
typical superconducting coherence length also varies a lot, ranging from 38 nm in Nb to 1600 nm in Al. Since the coherence 
length is defined as $\xi = \sqrt{D/\Delta}$ with the diffusion constant $D$ in the ferromagnetic region, we find that if $\varepsilon_T/\Delta = x$, then the thickness $d$ of the ferromagnetic region is given 
as $d = \xi/\sqrt{x}$. Assuming a superconducting material with $\xi=100$ nm, choosing $\varepsilon_T/\Delta = 0.1$ corresponds to 
$d \simeq 315$ nm. In order to make contact with the experimental situations that employ a ferromagnetic layer thickness of $d \simeq 20$ 
nm, we would then need a value of $x=25$. This is the motivation for our choice of the large Thouless energy $\varepsilon_T/\Delta=100$, 
corresponding to $d/\xi=0.1$. 
\par
Finally, we discuss the possible realization of a supercurrent-switch. The prospect of obtaining a dissipationless current-switching device relies on the opportunity to achieve sufficiently large residual values of the supercurrent at the 0-$\pi$ transition points. As seen from our results, this may be obtained for the intermediate barrier transparency regime, which agrees with the qualitative conclusions of Ref.~\onlinecite{zareyan2}. However, the effect of uniaxial spin-flip scattering has a detrimental effect on this residual value. While the width-dependence (Fig. \ref{fig:length}) shows some resilience towards an increased concentration of magnetic impurities, the 0-$\pi$ transition is highly sensitive to spin-flip scattering in temperature-dependence (Fig. \ref{fig:temp}). To switch the sign of the current, one has to increase either the width or temperature of the junction by an infinitesimal amount right at the transition point. Since only temperature can be manipulated in this way in a realisitic experiment, we arrive at the conclusion that the realization of a current-switch device in diffusive S/F/S junctions relies on samples with high quality interface and very low amounts of magnetic impurities.

\section{Summary}\label{sec:summary}
In conclusion, we have numerically studied the local density of states in a proximity ferromagnet/superconductor structure in the dirty limit. 
Our results take into account an arbitrary rate of spin-flip scattering and arbitrary interface transparency, focusing on a moderately transparent interface. This regime cannot be reached in the standard limiting cases of an ideal or tunneling interface. We have studied three 
cases for the size of the exchange field $h$ compared to the superconducting gap $\Delta$: \textit{i)} $h\lessim\Delta$, \textit{ii)} $h\gtrsim \Delta$, \textit{iii)} $h\gg \Delta$. In each case, we considered several values of the thickness of the ferromagnetic layer, barrier transparencies, and also different types of spin-dependent scattering to obtain the energy-resolved DOS. In doing so, we have clarified characteristic features that may be expected in the various parameter regimes accessible for the 
ferromagnetic film. Since our results take into account arbitrary proximity effect and magnetic impurity concentration, they should serve 
as a useful tool for a quantitative analysis of experimental data. In particular, we investigated the effect of spin-dependent scattering on the zero-energy behaviour observed in the DOS, which displayed the full range from a fully developed minigap to a peak-structure. By analyzing in detail the singlet and triplet part of the anomalous Green's function, we come to the important conclusion that it is necessary to distinguish between different types of spin-dependent scattering in order to correctly interpret DOS-measurements in ferromagnet/superconductor bilayers. Specifically, we find that the effect of spin-flip scattering (both uniaxial and isotropic) may differ fundamentally from the effect of spin-orbit scattering. The reason for this is that for weak exchange fields, the singlet component of the anomalous Green's function remains essentially unaltered by spin-orbit scattering, while the triplet component is strongly suppressed. 
\par
We have also investigated the supercurrent in diffusive superconductor/ferromagnet/superconductor junctions, allowing for arbitrary concentration of magnetic impurities and arbitrary interface transparency. We have investigated the effect of spin-flip scattering on the residual value of the supercurrent at the 0-$\pi$ transition points, and find a much weaker sensitivity to magnetic impurities in the width-dependence compared to the temperature-dependence of the $I_cR_N$-product. We have proposed that a finite and measurable residual value of the supercurrent may be obtained in the intermediate barrier transparency regime, although spin-flip scattering has a detrimental effect on this residual value. For samples with high quality interface and very low concentration of magnetic impurities, the residual value may be exploited to obtain efficient supercurrent-switching simply by altering the temperature of the system.

\acknowledgments
The authors gratefully acknowledge Y. Tanaka for useful discussions. J.L. and A.S. were supported by the Norwegian Research Council Grant Nos. 158518/431, 158547/431, (NANOMAT), 
and 167498/V30 (STORFORSK). T.Y. acknowledges support by the JSPS.

\appendix

\section*{Appendix}
\noindent We here define the star-product which enters the Eilenberger equation Eq. (\ref{eq:eilenberger}). For any two functions $A$ and $B$, we have
\begin{equation}
A\otimes B = \e{\i(\partial_{T_A}\partial_{\varepsilon_B} - \partial_{\varepsilon_A}\partial_{T_B})/2}AB,
\end{equation}
where the differentiation operators denote derivation with respect to the variables $T$ and $\varepsilon$ in the mixed representation. Note that if there is no explicit time-dependence in the problem, the star-product reduces to regular multiplication.
\par
The Pauli-matrices used in this paper are defined as \cite{jpdiplom}
\begin{align}
\underline{\tau_1} &= \begin{pmatrix}
0 & 1\\
1 & 0\\
\end{pmatrix},\;
\underline{\tau_2} = \begin{pmatrix}
0 & -\i\\
\i & 0\\
\end{pmatrix},\;
\underline{\tau_3} = \begin{pmatrix}
1& 0\\
0& -1\\
\end{pmatrix},\notag\\
\underline{1} &= \begin{pmatrix}
1 & 0\\
0 & 1\\
\end{pmatrix},\;
\hat{1} = \begin{pmatrix}
\underline{1} & \underline{0} \\
\underline{0} & \underline{1} \\
\end{pmatrix},\;
\hat{\tau}_i = \begin{pmatrix}
\underline{\tau_i} & \underline{0}\\
\underline{0} & \underline{\tau_i} \\
\end{pmatrix},\notag\\
\hat{\rho}_1 &= \begin{pmatrix}
\underline{0} & \underline{\tau_1}\\
\underline{\tau_1} & \underline{0} \\
\end{pmatrix},\;
\hat{\rho}_2 =  \begin{pmatrix}
\underline{0} & -\i\underline{\tau_1}\\
\i\underline{\tau_1} & \underline{0} \\
\end{pmatrix},\;
\hat{\rho}_3 = \begin{pmatrix}
\underline{1} & \underline{0}\\
\underline{0} & -\underline{1}  \\
\end{pmatrix}.
\end{align}

\end{document}